\documentclass[conference]{IEEEtran}
\IEEEoverridecommandlockouts

\usepackage{colortbl}
\usepackage{pifont}
\usepackage{booktabs} 
\usepackage{siunitx} 
\usepackage{multirow}

\usepackage{cite}
\usepackage{amsmath,amssymb,amsfonts}
\usepackage{algorithmic}
\usepackage{graphicx}
\usepackage{textcomp}
\usepackage{xcolor}
\def\BibTeX{{\rm B\kern-.05em{\sc i\kern-.025em b}\kern-.08em
    T\kern-.1667em\lower.7ex\hbox{E}\kern-.125emX}}
\begin{document}

\title{Toward Adaptive BCIs: Enhancing Decoding Stability via User State-Aware EEG Filtering\\
\thanks{This work was partly supported by Institute of Information \& Communications Technology Planning \& Evaluation (IITP) grant funded by the Korea government (MSIT) (No. RS-2019-II190079, Artificial Intelligence Graduate School Program (Korea University) and No. RS-2021-II-212068, Artificial Intelligence Innovation Hub).}
}

\author{\IEEEauthorblockN{Yeon-Woo Choi}
\IEEEauthorblockA{\textit{Dept. of Artificial Intelligence} \\
\textit{Korea University}\\
Seoul, Republic of Korea \\
yw\_choi@korea.ac.kr}
\and
\IEEEauthorblockN{Hye-Bin Shin}
\IEEEauthorblockA{\textit{Dept. of Brain and Cognitive Engineering} \\
\textit{Korea University}\\
Seoul, Republic of Korea \\
hb\_shin@korea.ac.kr}
\and
\IEEEauthorblockN{Dan Li}
\IEEEauthorblockA{\textit{Dept. of Artificial Intelligence} \\
\textit{Korea University}\\
Seoul, Republic of Korea \\
dan\_li@korea.ac.kr}
\and
}

\maketitle

\begin{abstract}
Brain-computer interfaces (BCIs) often suffer from limited robustness and poor long-term adaptability. Model performance rapidly degrades when user attention fluctuates, brain states shift over time, or irregular artifacts appear during interaction. To mitigate these issues, we introduce a user state-aware electroencephalogram (EEG) filtering framework that refines neural representations before decoding user intentions. The proposed method continuously estimates the user’s cognitive state (e.g., focus or distraction) from EEG features and filters unreliable segments by applying adaptive weighting based on the estimated attention level. This filtering stage suppresses noisy or out-of-focus epochs, thereby reducing distributional drift and improving the consistency of subsequent decoding. Experiments on multiple EEG datasets that emulate real BCI scenarios demonstrate that the proposed state-aware filtering enhances classification accuracy and stability across different user states and sessions compared with conventional preprocessing pipelines. These findings highlight that leveraging brain-derived state information—even without additional user labels—can substantially improve the reliability of practical EEG-based BCIs.
\end{abstract}


\section{INTRODUCTION}
Brain–computer interfaces (BCIs) are designed to bridge human brain and machine execution, enabling seamless interaction between users and computational systems \cite{Wolpaw2002bci}. For such systems to be practically viable, they should be capable of accurately decoding user intentions and maintaining stable performance over time. However, BCIs continue to face persistent obstacles that limit their long-term adaptability \cite{HMIs}.
A critical issue stems from the dynamic and non-stationary nature of human users.
The user's mental and physiological states naturally fluctuate over time, which alters data distributions and introduces outlier samples or transient noise into the data stream \cite{ourlab_1}. This compromises feature representations and degrades decoding reliability. Even well-trained models can experience severe degradation when these internal user states shift. 
In practice, prior studies across various modalities—including ambient environment, hand-drawn trajectories, and cognitive states—have emphasized the importance of identifying data consistency or leveraging it for model training \cite{ourlab_2, ourlab_3, Kaushik_AS, ourlab_11, Myrden_AS}.
In particular, adapting to changing outdoor conditions is crucial \cite{ourlab_7}, as environmental variations—such as illumination, weather, or background dynamics—and internal fluctuations in user attention or cognitive engagement can significantly influence signal patterns and model performance \cite{state1}.

To address this issue, researchers have increasingly turned toward brain-derived user-state monitoring as a means of contextual adaptation.
Various studies have demonstrated that neurophysiological signals such as electroencephalogram (EEG) can be used to infer user states including emotion, concentration, mental workload, and alertness \cite{ourlab_4, b19, b20, ourlab_13, RomeroAS, Komarov_AS, ourlab_15}.
Beyond short-term fluctuations in cognitive states, EEG has also been utilized to identify more persistent neurological conditions or disease-related patterns.
For example, Prabhakar \emph{et al.}~\cite{ourlab3} developed an EEG-based schizophrenia classification framework using nature-inspired optimization techniques to extract discriminative neural features, showing that EEG carries rich information about underlying neural and cognitive states.
Moreover, strong correlations have been observed between such user states and the controllability or accuracy of BCIs, emphasizing that user attention is a key determinant of decoding performance \cite{RiddleAS, KnyazevAS, WardAS}.
Several methods have attempted to exploit this relationship by modulating the decoder output—for example, adjusting classifier thresholds or control gains according to estimated attention or fatigue indices \cite{state2, state3}.
While these approaches have proven useful for stabilizing system output, they generally operate at the decision level, leaving the underlying representation of neural data unchanged.

In this work, we take a different approach and propose a user state-aware EEG filtering framework that directly enhances the quality of neural features before decoding.
The key idea is to estimate the user’s cognitive engagement level from EEG signals and selectively retain EEG segments that correspond to attentive mental states.
EEG signals associated with transient distraction are discarded, whereas highly focused epochs are preserved to guide more reliable decoding.
Moreover, beyond data-level filtering, various deep learning architectures have been developed to achieve robust decoding under noisy, non-stationary, and context-dependent conditions \cite{deeplearning, ourlab6}.
In particular, the first attempt to achieve adaptive learning at the model level was introduced by Bengio \emph{et al.}~\cite{first_CL}, 
where training samples were re-weighted based on their sample-wise loss values, effectively filtering out unreliable samples at each iteration.
This hierarchical adaptation, encompassing both signal- and model-level filtering, establishes a foundation for stable decoding under real-world, non-stationary EEG dynamics.


\begin{figure}[!t] 
\centering
\includegraphics[width=\columnwidth]{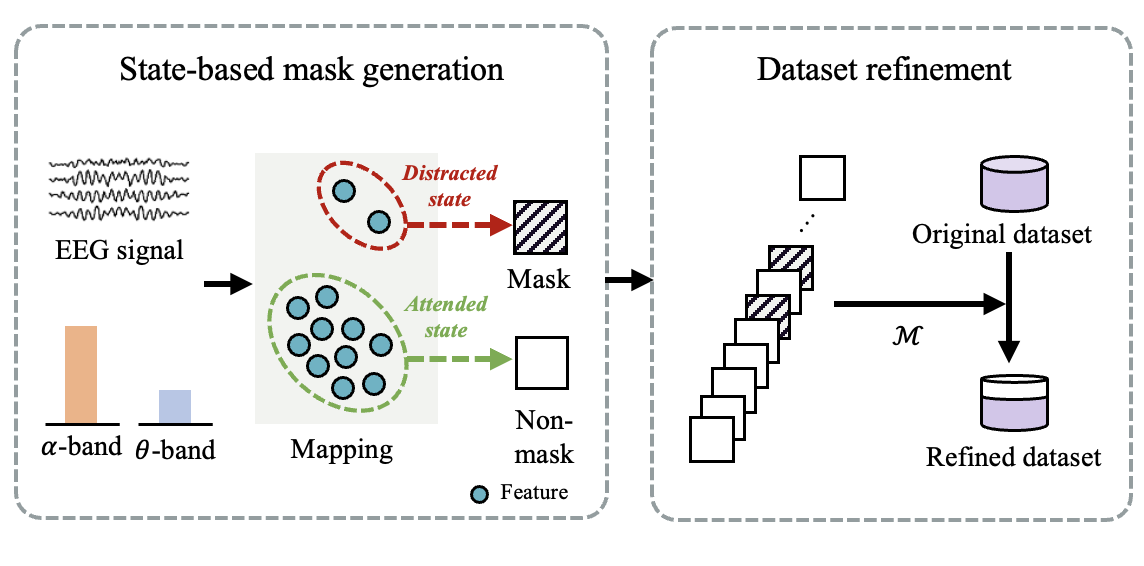} \vspace{-0.6cm}
\caption{Illustration of the user state-based binary masking process for robust decoding against state fluctuations. Energy from attention-related frequency bands is extracted from EEG signals and mapped to feature space to generate masks, which are then applied to the existing training dataset for refinement.}
\label{fig2}
\end{figure}

\section{METHODS}
\subsection{Overview}
The proposed framework combines user-state estimation and adaptive sample selection to stabilize EEG decoding. Alpha–Theta spectral ratios are used to identify and remove distracted EEG segments, yielding attention-consistent training data, as shown in Fig. 1.
During training, sample-wise loss filtering further suppresses unstable or high-loss trials, enabling gradual adaptation to more complex data.
Together, these dual-stage filtering mechanisms enhance decoding robustness and cross-session generalization across backbone models.

\subsection{EEG Preprocessing and Feature Extraction}
EEG signals were band-pass filtered between 1--40~Hz using a fourth-order
Butterworth filter and down-sampled to 250 Hz.
Each trial was segmented into 2 s non-overlapping windows and standardized
per channel.
Spectral features were obtained by computing the power spectral density
(PSD) in canonical frequency bands. These band powers served as input features for both
user-state estimation and subsequent classification.

\subsection{State–Aware Filtering}
EEG rhythms in the Alpha (8–13 Hz) and Theta (4–8 Hz) bands are closely associated with user attention levels: elevated Alpha power reflects reduced engagement, whereas enhanced Theta power indicates sustained concentration \cite{RiddleAS, WardAS}. 
To quantify this relationship, we compute the Alpha–Theta ratio (ATr) by estimating the spectral power in each frequency band via fast Fourier transform (FFT), 
which has been widely used for analyzing EEG signals \cite{ourlab4}.
To quantify attentional variation, we compute an \textit{attention index} for each trial as:
\begin{equation}
ATr_i = \frac{E_{\alpha}(i)}{E_{\theta}(i)},
\end{equation}
where $E_{\alpha}$ and $E_{\theta}$ denote band-specific spectral energies obtained through FFT analysis. 
Samples with abnormally high $ATr$ values, corresponding to strong distraction or transient artifacts, are identified using Tukey’s outlier criterion:
\begin{equation}
M(s_i) =
\begin{cases}
1, & s_i \le Q_3 + k(Q_3 - Q_1),\\
0, & \text{otherwise.}
\end{cases}
\end{equation}

Here, $Q_1$ and $Q_3$ represent the quartiles of the $ATr$ distribution, $s_i$ denotes the $i$-th trial’s attention ratio value derived from the corresponding EEG segment, and $k$ is a subject-specific threshold. 
To ensure optimal subject-specific sensitivity, the value of $k$ was fine-tuned individually for each participant during the training phase.
Specifically, classification performance was evaluated across a predefined range of $k$ values, and the parameter yielding the highest accuracy was retained for that subject.
The selected $k$ was then consistently applied to subsequent sessions to maintain cross-session comparability.
Once the optimal threshold was determined, the corresponding mask $M(s_i)$ was applied to exclude segments identified as outliers, thereby refining the dataset to include only attention-consistent EEG epochs.
This filtering step mitigates session variability and reduces the negative influence of cognitively inconsistent EEG patterns.

After signal-level refinement, model training further incorporates sample-wise loss-based filtering, 
following the adaptive learning mechanism described in \cite{JiangCLsurvey}. 
For each input $x_i$ with label $y_i$, an individual loss is computed as:
\begin{equation}
L_i = \mathcal{L}(y_i, f(x_i; \theta)),
\end{equation}
where $f(x_i; \theta)$ denotes the decoder prediction and $\mathcal{L}$ represents the cross-entropy loss. 
Samples producing excessively high $L_i$ are considered unreliable or inconsistent and are temporarily excluded from the active training pool: 
\begin{equation}
w_i =
\begin{cases}
1, & L_i < \lambda,\\
0, & \text{otherwise,}
\end{cases}
\end{equation}
where $\lambda$ is a gradually increasing threshold controlling the inclusion of more difficult samples as training progresses. 
This process functions as a dynamic sample filter that prioritizes low-loss, high-confidence EEG trials in the early stage and incrementally introduces more variable samples as the model stabilizes. 

\subsection{Datasets}
Motor imagery (MI)–based BCI paradigms are widely used benchmarks for evaluating EEG decoding algorithms because they provide multiple well-controlled datasets with standardized acquisition protocols \cite{MI, ourlab_14}.
To verify the effectiveness of the proposed user state-aware filtering framework, we evaluated it using three representative MI datasets provided by the Mother of All BCI Benchmarks (MOABB) platform \cite{moabb}.
All datasets involve multi-session recordings, allowing us to examine the robustness of the filtering scheme under varying user states and temporal conditions.

\begin{figure}[!t] 
\centering
\includegraphics[width=\columnwidth]{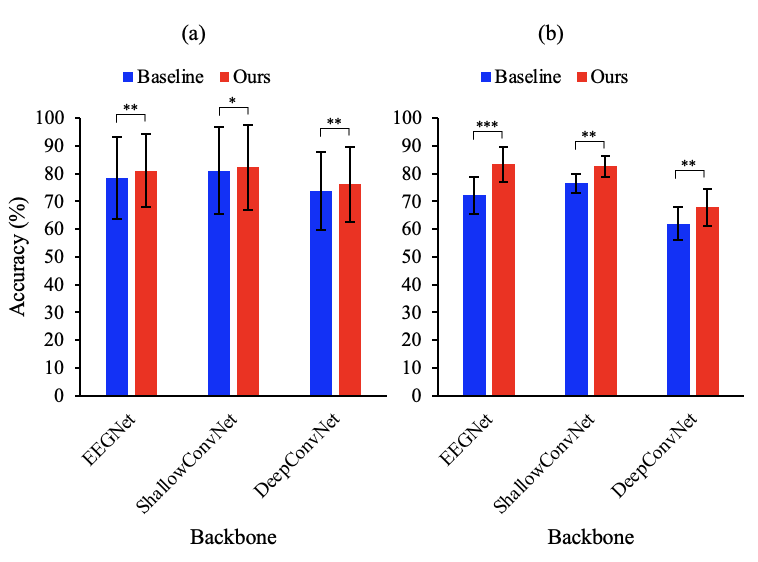} \vspace{-0.5cm}
\caption{Performance comparison between baseline and proposed methods across various network architectures on (a) BNCI2014004 and (b) Zhou2016 datasets. Stars denote statistical significance between baseline and our proposed method: $^*$\(\textit{p} < 0.05\), $^{**}$ \(\textit{p} < 0.01\), $^{***}$ \(\textit{p} < 0.001\).}
\label{fig2}
\end{figure}

BNCI2014004 dataset corresponds to the well-known BCI Competition IV-2b benchmark \cite{bcicomp2b}.
Nine participants performed left- and right-hand motor imagery tasks across five sessions.
To exclude potential feedback effects, we used the first two sessions that did not include visual feedback.
Each session contained 120 trials, and EEG signals were recorded from three electrodes (C3, Cz, and C4) at a sampling rate of 250 Hz.
Signals were low-pass filtered at 38 Hz using a fourth-order Butterworth filter.
This dataset serves as a compact benchmark for evaluating cross-session consistency under minimal spatial coverage.

To evaluate the model’s temporal robustness, we used the Zhou2016 dataset \cite{zhou} from the MOABB repository, which provides multi-session EEG recordings.
The dataset includes four participants who performed three motor imagery tasks—imagining movements of the left hand, right hand, and both feet—over three separate recording sessions.
Each session contains 160 trials collected from 14 EEG channels at a 250 Hz sampling rate.
The time gap between consecutive sessions ranges from several days to a few months, offering a suitable testbed for assessing cross-session stability and long-term adaptability.

\section{RESULTS AND DISCUSSION}

\subsection{Classification Performance}
Fig. 2(a) shows the classification performance of our proposed method applied to representative backbone models—EEGNet, ShallowConvNet, and DeepConvNet—on the BNCI2014004 dataset, which have been the most widely adopted architectures following the traditional approach \cite{ourlab_8, ourlab_12} in biosignal classification research. Compared to baseline, our approach demonstrated significantly superior average performance across participants, with accuracy improvements of 2.72 \%, 1.28 \%, and 2.30 \% for each backbone model, respectively. The reduction in standard deviation also indicates improved consistency across subjects. For the Zhou2016 dataset, shown in Fig. 2(b), even larger gains were observed: average accuracy improvements of 11.17 \% for EEGNet, 6.08 \% for ShallowConvNet, and 5.82 \% for DeepConvNet, all showing statistically significant gains over the baseline models. The reduced standard deviations across subjects further indicate enhanced inter-participant consistency and decoding stability. Among the three architectures, ShallowConvNet achieved the highest overall performance on both datasets, reflecting its strong ability to extract mid-level spatio-spectral patterns that are well aligned with the attention-filtered EEG representations. 

\begin{figure}[!t] 
\centering
\includegraphics[width=\columnwidth]{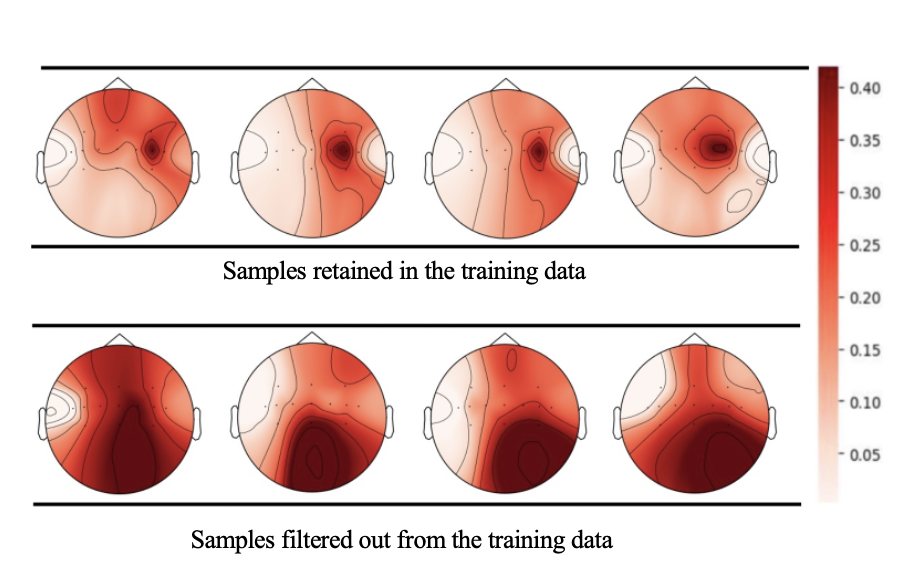} \vspace{-0.4cm}
\caption{Alpha PSD topographical maps demonstrate localized patterns characteristic of motor imagery task in included samples, while excluded samples exhibit elevated occipital alpha power indicative of user distraction, with all visualizations derived from the Zhou2016 dataset.}
\label{fig2}
\end{figure}

Overall, the results confirm that integrating attention-based outlier removal with confidence-guided sample selection leads to cleaner EEG representations and enhances model robustness. By prioritizing cognitively stable, high-confidence samples, the proposed filtering pipeline effectively stabilizes decoding across architectures and datasets, moving toward more adaptive and reliable BCI systems.

\subsection{Neurophysiological Analysis}
To examine the neurophysiological relevance of the user-state-based filtering, we analyzed whether EEG segments categorized as high-quality or low-quality displayed distinct spectral or spatial features. As illustrated in Fig. 3, the Alpha band power PSD topographies were compared between the samples retained and those removed by the proposed filtering process for a representative participant from the Zhou2016 dataset. The upper row represents EEG samples retained as high-quality, corresponding to trials with low decoding error that were included in the early stage of model learning. These samples exhibited well-localized neural activation, particularly around the motor cortical area—an activation pattern commonly associated with focused motor imagery execution and sustained attention. Conversely, the lower row displays the samples excluded by the filtering step, which showed pronounced Alpha activity over the occipital region. This spatial pattern indicates reduced attentional engagement or partial disengagement during task performance, suggesting that the filtering effectively distinguishes between attentive and distracted states in the neurophysiological domain.

\section{CONCLUSIONS}
This work presented a user state-aware EEG filtering framework to enhance the stability and adaptability of brain–computer interfaces.
By integrating cognitive-state estimation into preprocessing, the framework filters out distracted or inconsistent EEG segments while preserving reliable neural representations.
Experiments across multiple datasets and architectures demonstrated significant gains in decoding accuracy and inter-subject stability.
Operating at the signal-representation stage, our method mitigates distributional drift prior to training and aligns learning with reliable samples, improving robustness under fluctuating user conditions.
These findings highlight the potential of incorporating intrinsic brain-state information for building more adaptive and attention-aware BCIs.
Future work will focus on real-time state monitoring and integration with closed-loop adaptive decoders to ensure long-term reliability in practical human–machine interaction and neurorehabilitation scenarios.

\end{document}